\documentclass[twocolumn,showpacs,preprintnumbers,amsmath,amssymb]{revtex4}
\usepackage{tabularx,graphicx}\begin{document}
\newcommand{\beq}{\begin{equation}}
\newcommand{\eeq}{\end{equation}}
\newcommand{\beqn}{\begin{eqnarray}}
\newcommand{\eeqn}{\end{eqnarray}}
\newcommand{\bmath}{\begin{subequations}}
\newcommand{\emath}{\end{subequations}}
\newcommand{\bra}[1]{\langle #1|}
\newcommand{\ket}[1]{|#1\rangle}

\title{Experimental consequences of  predicted charge rigidity of superconductors}
\author{J. E. Hirsch }
\address{Department of Physics, University of California, San Diego,
La Jolla, CA 92093-0319}

\begin{abstract} 
The theory of hole superconductivity predicts that in superconductors the charged superfluid is about a million times more rigid than the normal electron fluid. 
We  point out that this physics should give rise to large changes in
the bulk and surface plasmon dispersion relations of metals entering the superconducting state, that have not yet been experimentally detected and would
be in stark contradiction with the expected behavior 
within conventional BCS-London theory.
We  also propose that this explains the
puzzling experimental observations of Avramenko et al\cite{sound} on electron sound propagation in superconductors and the puzzling experiments of W. de Heer et al\cite{clusters} 
detecting large electric dipole moments in small metal clusters, as well as the Tao effect\cite{tao} on aggregation of superconducting microparticles in an electric field. 
Associated with the enhanced charge rigidity is   a large increase in the electric screening length of superconductors at low temperatures that has not yet been experimentally detected.
The physical origin of the enhanced charge rigidity and its relation to other aspects of the theory of hole superconductivity is discussed.
\end{abstract}
\maketitle 

\section{introduction}
A normal metal screens electrostatic fields over distances of order of the interelectronic distance, or $k_F^{-1}$, with $k_F$ the Fermi wavevector, quantitatively of order $\AA^{-1}$
for normal metallic densities. 
Conventional BCS-London theory predicts that the response of superconductors to a static electric field is essentially the same as that of  normal
metals\cite{conv0,conv1,conv1p,conv2,conv3,conv4}. Instead, the theory of hole superconductivity\cite{holetheory} predicts that superconductors can only screen electrostatic fields over much larger distances, in their ground state over
distances of order  $\lambda_L$, the London penetration depth, quantitatively of order hundreds of $\AA$\cite{chargeexp}.

Within the theory of hole superconductivity the inability of superconductors to screen over smaller distances
originates in the fact that superconducting  electrons reside in highly overlapping orbits of radius $2\lambda_L$\cite{sm}, and this  precludes the possibility of charge fluctuations of shorter
wavelengths that would destroy the ability of the superconducting electrons to maintain phase coherence as they traverse these large orbits. That superfluid electrons
reside in orbits of radius $2\lambda_L$ follows from the fact that according to this theory this is the only consistent way to explain $dynamically$ the Meissner effect
exhibited by all superconductors\cite{lastpaper}. The inability of superfluid electrons to screen leads, via the compressibility sum rule, to the prediction that superfluid electrons are highly 
incompressible unlike normal metal electrons.

This enhanced rigidity of the superconducting fluid implies that the longitudinal plasmon dispersion relation will  be much steeper in the
superconducting than in the normal state\cite{electrodyn}. In contrast, BCS theory predicts no change in the plasmon dispersion relation\cite{conv1,conv1p,conv2,conv3}.
The bulk plasmon dispersion relation can be measured  by EELS (electron energy loss spectroscopy)\cite{raether,eels} and by  inelastic X-ray scattering\cite{xrays5}
as well as
optically in transmission experiments through thin films\cite{ploptical0,ploptical}. Also the surface plasmon dispersion relation, which can be measured
in EELS\cite{surfeels}  or optical experiments\cite{surfacepl}, should  change in the superconducting state. To our knowledge these experiments have not been yet done
on superconductors. We discuss here what we expect the observations will show, in stark contrast with what would be expected within BCS-London theory.

We furthermore discuss three experiments that have been performed in recent years that provide strong evidence in favor of the enhanced rigidity of 
superfluid electrons predicted by our theory: (i) sound propagation by electrons (Avramenko effect)\cite{sound}, (ii) electric dipole moments of small 
metal clusters (de Heer effect)\cite{clusters},
and (iii) aggregation of superconducting microparticles in large electric fields (Tao effect)\cite{tao}.

The larger electric screening length of superconductors should be directly detectable experimentally. So far the only experimental indication of this  appears to be a report
  by   Jenks and Testardi\cite{testardi} that measured an increased
penetration of electric field in $YBCO$ films below $T_c$. 
We discuss the expected behavior of the electric screening length below $T_c$ within our theory.

\section{electrodynamic equations for superconductor}
Within the theory of hole superconductivity the first London equation for the time derivative of the supercurrent is modified to read\cite{electrodyn}
\beq
\frac{\partial \vec{J}_s}{\partial t}=\frac{n_s e^2}{m_e}(\vec{E}+\vec{\nabla}\phi)
\eeq
with $\phi$ the electric potential (the $\vec{\nabla}\phi$ -term is absent in the conventional London equations). The magnetic vector potential $\vec{A}$ in the second London
equation
\beq
\vec{J}_s=-\frac{c}{4\pi \lambda_L^2}\vec{A}
\eeq
obeys the Lorenz gauge $\vec{\nabla}\cdot\vec{A}=-(1/c)(\partial \phi/\partial t)$ rather than the London gauge $\vec{\nabla}\cdot\vec{A}=0$\cite{electrodyn}. Note that Eq. (1) follows from
Eq. (2) on using Faraday's law. The charge density in the superconductor $\rho(\vec{r},t)$ satisfies the equation\cite{electrodyn}
\beq
\frac{\partial^2 \rho}{\partial t^2} +\frac{c^2}{\lambda_L^2}(\rho-\rho_0)=c^2 \nabla^2\rho 
\eeq
and the electric potential $\phi(\vec{r},t)$ satisfies the same equation
\beq
\frac{\partial^2 \phi}{\partial t^2} +\frac{c^2}{\lambda_L^2}(\phi-\phi_0)=c^2 \nabla^2\phi
\eeq
where $\rho_0$ is a uniform positive charge density and $\phi_0(\vec{r})$ is the resulting electrostatic potential ($\nabla^2 \phi_0=-4\pi\rho_0$). The London
penetration depth $\lambda_L$ is given by the usual form\cite{tinkham}
\beq
\frac{1}{\lambda_L^2}=\frac{4\pi n  e^2}{m_e c^2}=\frac{\omega_p^2}{c^2}
\eeq
with $\omega_p$ the plasma frequency. The parameter $\rho_0$ is determined by the condition that the internal electric field 
that develops in the interior of the superconductor due to expulsion of negative charge to the surface\cite{chargeexp} should reach its maximum value\cite{electrospin}
\beq
E_m=-\frac{\hbar c}{4e\lambda_L^2}
\eeq
within a London penetration depth from the surface,
pointing outward perpendicular to the surface.

\section{dielectric function and compressibility}
It follows from the electrodynamic equations discussed in the previous section that the longitudinal dielectric function for the 
superfluid   within our theory  is given  by\cite{electrodyn}
\beq
\epsilon_s(k,\omega)=1-\frac{\omega_p^2}{\omega^2-c^2k^2}  .
\eeq
Eq. (7) is of the generic form of a hydrodynamic longitudinal dielectric function for the electron fluid\cite{hydro1,hydro2}
\beq
\epsilon_l (k,\omega)=1-\frac{\omega_p^2}{\omega^2-\beta^2 k^2}
\eeq
that yields for the static dielectric constant
\beq
\epsilon_l (k,\omega\rightarrow 0)=1+\frac{\omega_p^2}{\beta^2 k^2}= 1+\frac{4\pi e^2 n^2 \kappa}{k^2}.
\eeq
with $\kappa$ the electronic compressibility.
The second equality follows from the compressibility sum rule\cite{csr}, so that 
\beq
\beta^2=\frac{1}{m_e n\kappa}
\eeq
with \beq
\kappa=-\frac{1}{V}\frac{\partial V}{\partial P}   .
\eeq
.

For the free electron gas, the zero temperature compressibility is 
\beq
\kappa = \frac{3}{2n\epsilon_F}=\frac{3}{nm_e v_F^2}  ,
\eeq
with $\epsilon_F=m_e v_F^2/2$ the Fermi energy and $v_F$ the Fermi velocity,  
yielding 
\beq
\beta^2=\frac{1}{3} v_F^2 .
\eeq
Instead, for the superconductor we have from Eqs. (7) and (8)
\beq
\beta^2=c^2
\eeq
\beq
\kappa_s=\frac{1}{n_s m_e c^2}
\eeq
so that the superconducting electron fluid is enormously  more rigid than the normal metal electron fluid, since $c>>v_F$.
We expect this enhanced rigidity to show up in experiments where electron density oscillations are induced that are
$not$ accompanied by motion of the ions so that an electric potential builds up in the interior of the superconductor.

Eq. (9) yields  the static longitudinal dielectric functions for the superfluid electrons  and the normal metal electrons respectively
\beq
\epsilon_s(k,0)=1+\frac{1}{\lambda_L^2 k^2}
\eeq
\beq
\epsilon_n(k,0)=1+\frac{1}{\lambda_{TF}^2 k^2}
\eeq
with $\lambda_L$ the London penetration depth given by Eq. (5) and $\lambda_{TF}$ the Thomas Fermi screening length given by
\beq
\frac{1}{\lambda_{TF}^2}=\frac{6\pi n  e^2}{\epsilon_F}=\frac{4}{\pi a_0 k_F^{-1}}
\eeq
with $a_0$ the Bohr radius.

\section{bulk  plasmons}
The bulk plasmon dispersion relation follows from setting the longitudinal dielectric function to zero. Eq. (8) yields
\beq
\omega_k^2=\omega_p^2+\beta^2 k^2
\eeq
hence we predict for superconductors at zero temperature
\beq
\omega_k^2=\omega_p^2+c^2 k^2
\eeq
For the normal metal, the plasmon dispersion relation obtained from the longitudinal dielectric function calculated in the random phase approximation
(Linhardt dielectric function) yields Eq. (19) with\cite{linhardt}
\beq
\beta^2=\frac{3}{5} v_F^2
\eeq
which is slightly different from Eq. (13), valid in the low frequency limit\cite{sham}. Thus, to reproduce the Linhardt dielectric function the hydrodynamic form
Eq. (8) has to include a variation of $\beta^2$ from low to high frequencies.

Within a two-fluid description of a metal below the superconducting transition temperature the electronic pressure results from the sum of the superfluid and the
normal fluid pressures, hence
\beq
\frac{1}{\kappa}=\frac{1}{\kappa_s}+\frac{1}{\kappa_n}
\eeq
and the parameter $\beta^2$ is
\beq
\beta^2(T)=\frac{1}{nm_e}(\frac{1}{\kappa_n}+\frac{1}{\kappa_s})=  \frac{n_n}{n}\frac{3}{5} v_F^2+\frac{n_s}{n} c^2
\eeq
where $n_n$ and $n_s$ are the densities of normal and superfluid electrons and we have used the high frequency value of $\beta$ Eq. (21) for the normal 
electron contribution.

Thus, Eqs. (19) and (23) show that as the temperature is lowered below $T_c$ a sharp increase in the slope of the bulk plasmon dispersion relation 
$\omega_k^2$ versus $k^2$ should be seen. In a two-fluid model one has  $n_s=n(1-t^4), n_n=nt^4$, with $t=T/T_c$, 
which is also approximately the behavior predicted by BCS theory\cite{tinkham}, so 
we expect for the bulk plasmon dispersion relation
\beq
\omega_k^2=\omega_p^2+[t^4\frac{3}{5}v_F^2 + (1-t^4)c^2]k^2
\eeq

   \begin{figure}
 \resizebox{8.5cm}{!}{\includegraphics[width=9cm]{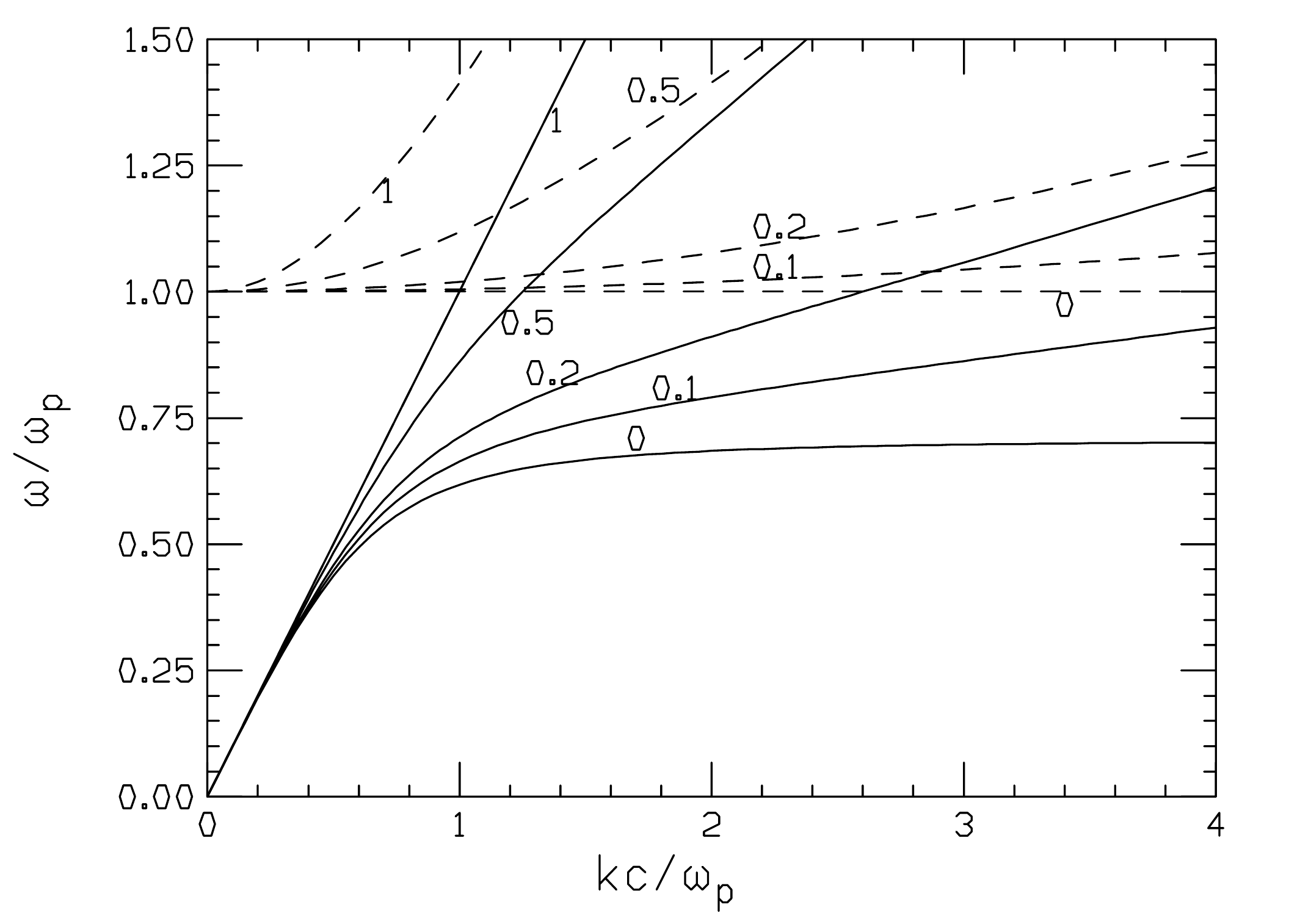}}
 \caption {Surface (solid lines) and bulk (dashed lines) plasmon dispersion relations for different values of $\beta/c$ (numbers next to the lines).
 $\beta$ increases as the temperature is lowered (Eq. 29)).
 The values of $\beta/c$ of $0, 0.1, 0.2$ and $0.5$  correspond to values of $T/T_c$ of $1$, $0.998$, $0.990$ and $0.931$ respectively.
 }
 \label{figure2}
 \end{figure}
 
 \section{surface plasmons}
 Surface plasmons (also called surface plasmon polaritons) are  longitudinal charge oscillations   coupled to  an electromagnetic wave
 with both longitudinal and transverse  field components  propagating along the surface of a metal, excited by either
 fast electrons or electromagnetic radiation.
Crowell and Ritchie\cite{surfpl} and Fuchs and Kliewer\cite{hydro2} derived the following dispersion relation\cite{surfpl}
 \beqn
 &&\omega_k^2[(k^2c^2+\omega_p^2-\omega_k^2)^{1/2}+(k^2c^2-\omega_k^2)^{1/2}]\times  \nonumber \\
 &&(k^2\beta^2+\omega_p^2-\omega_k^2)^{1/2}= \nonumber \\
 & &\omega_p^2[k^2 c \beta+(k^2c^2-\omega_k^2)^{1/2}(k^2\beta^2+\omega_p^2-\omega_k^2)^{1/2}]     .
 \eeqn
 In the limit  $\beta\rightarrow 0$ the solution is
 \beq
 \omega_k^2=\frac{\omega_p^2}{2}+k^2 c^2 -\sqrt{\frac{\omega_p^4}{4}+k^4c^4}
 \eeq
 and $\omega_k\rightarrow \omega_p/\sqrt{2}$ for large $k$. For any $\beta\neq 0$, the surface plasmon dispersion relation for large $k$ is, from Eq. (25)
 \beq
 \omega_k^2=\omega_p^2+\beta^2k^2-\frac{\omega^4}{4\beta^2(1-\frac{\beta^2}{c^2})k^2}
 \eeq
 so it approaches the bulk plasmon dispersion relation Eq. (19). 
 For small wavevectors the surface plasmon dispersion relation is
 \beq
 \omega_k^2= c^2k^2-\frac{(1-\frac{\beta}{c})^2}{\omega_p^2}k^4c^4
 \eeq
so it increasingly deviates from the transverse dispersion relation $\omega_k=ck$ for smaller $\beta$ and larger $k$.
 
Figure 1 shows examples of the surface and bulk plasmon dispersion relations for various $\beta$. As function of temperature, $\beta$ increases very rapidly as
$T$ is lowered below $T_c$ according to Eq. (23). For any typical value of $v_F$ (of order $1\%$ of the speed of light) the $v_F$ term in Eq. (23) can be ignored, so that
\beq
\beta^2(T)=(1-t^4)c^2 .
\eeq
The values of $\beta$ of $0, 0.1, 0.2$ and $0.5$ shown in Fig. 1 correspond to values of $T/T_c$ of $1$, $0.998$, $0.990$ and $0.931$ respectively.
Consequently we expect   rapid changes in the observed surface and bulk plasmon frequencies as the temperature is lowered below $T_c$, in contrast to conventional
BCS-London theory that predicts no change\cite{conv3}. 

However, for surface plasmons the interpretation of experiments could be  more complicated 
because it appears that in experiments performed in the normal state
the induced charge fluctuations can spill out of the surface, drastically modifying the 
dispersion relation\cite{plummer,feibelman,bennett}, an effect which is not taken into account by Eq. (25). We expect this spill-out effect 
to be even more pronounced in the superconducting state because of the enhanced rigidity and because the superconductor has an enhanced tendency to 
spill out electrons within our theory\cite{giantatom}.

\section{plasmon  experiments}
Measurement of the angular dependence of scattered electrons in electron energy loss experiments (EELS) provides information on the bulk plasmon dispersion relation. A large number of such studies has been performed on many different normal metals since the 1950's\cite{raether}. 
Generally these studies are done at room temperature, although there have also been EELS studies of the effect of temperature on the plasma frequency
down to liquid helium temperatures for $Al$\cite{allowt} and $Pb$\cite{pblowt}. However to our knowledge there has not been  a single EELS study of plasmons in a superconducting metal
that would look at possible changes in the plasmon dispersion relation below the critical temperature (except for ref. \cite{fink} for a high $T_c$ cuprate
that did not detect any change presumably due
to experimental accuracy limitations). 
This is very surprising and we hope such experiments will be done in the near future. As discussed in Sect. IV we expect a very rapid increase in the
plasmon energy for fixed wavevector as the system is cooled below $T_c$.

Longitudinal bulk plasmons can also be excited optically with obliquely incident p-polarized (parallel to the plane of incidence) electromagnetic radiation\cite{ploptical0,ploptical}
and the plasmon dispersion relation can be measured. For example, Lindau and Nilsson\cite{ploptical} 
obtained the bulk plasmon $\omega_k$ for $Ag$ from transmission experiments through thin  films of different thicknesses of the order $100\AA$. The experiment is done at fixed angle of
incidence and each film thickness gives a small number (2 in this case) of points in  the dispersion relation.  Anderegg et al\cite{anderegg} measured the plasmon
dispersion relation for  $K$ from oscillatory structure in the absorption of thin films of varying thickness from $27\AA$ to $100\AA$, and were able to extract up to
$10$ data points per film.  Again, it would seem straightforward to do such experiments  with superconducting films but not a single study has been performed so far
to our knowledge. We hope such studies will be done in the near future.

Inelastic X-ray scattering (IXS) experiments can also provide information on the bulk plasmon dispersion relation\cite{xrays,xrays2,xrays4,xrays3}. 
IXS experiments at low temperatures have been performed in recent years for   example to study the physics of liquid and solid $He$\cite{xrayshe0,xrayshe,xrayshe2}. 
However no attempt has been made to date to
study the plasmon dispersion relation of metals like e.g. $Al$\cite{xrays4} in the temperature range where they become superconducting using this technique.

Surface plasmon experiments on superconductors have never been performed to our knowledge, neither EELS nor IXS nor optical.  With conventional optical methods it is complicated to 
excite surface plasmons  because of required matching conditions  and rough surfaces are needed which introduces
additional complications\cite{raethersurface}. However, recently developed scanning near-field optical microscopy
techniques\cite{basovreview} provide the possibility to locally excite and detect surface plasmons\cite{basov} and may  allow for detailed studies of
the effect of the onset of superconductivity on the surface plasmon dispersion relation.

Finally, surface plasmons excited in metal nanoparticles (Mie resonances)\cite{mie} are sensitive to the longitudinal dielectric function\cite{mie2} and thus are 
likely to  show interesting changes due to the change in the dielectric response  that
we predict upon onset of superconductivity. Such experiments have 
never been done with superconducting nanoparticles to our knowledge.

\section{Electron sound anomaly}
Avramenko and coworkers\cite{sound} apply a longitudinal elastic wave to the surface of a metal and detect an electric potential oscillation at the opposite end of
the sample. They find two types of signals, one propagating at the ordinary sound velocity and a much faster one propagating at a speed of order the Fermi velocity, which they 
call ``electron sound''. When the temperature of the sample is lowered below the superconducting transition temperature the amplitude of the transmitted signals drops
precipitously. Avramenko et al point out that this behavior has no explanation within the conventional theory of superconductivity.

According to Avramenko et al the displacement amplitude at the receiving interface for the electron sound signal is
\beq
u_{ES} \sim  \frac{s}{v_{eff}}u_0
\eeq
where $s$ is the sound velocity and $v_{eff}$ the velocity of electron sound propagation which Avramenko et al assume is the Fermi velocity $v_F$. 
$u_0$ is  the amplitude of the elastic vibrations at the interface where the signal is generated.  $u_{ES}$ determines the electric potentials measured at the
receiving interface $\varphi_S$ and $\varphi_{ES}$ for sound and electron sound. Both potentials decrease precipitously as the temperature is lowered below the
superconducting $T_c$.

If the superfluid is very rigid compared to the normal fluid as predicted by our theory (Eq. (15)) it is natural to expect that the amplitude of longitudinal charge oscillations will rapidly decrease
as the  temperature is lowered below $T_c$ and the
superfluid concentration increases. Following the behavior of the bulk modulus Eq. (23) we  argue that the electron sound velocity $v_{eff}$ in Eq. (30)   below $T_c$ 
can be estimated by
\beq
v_{eff}=\sqrt{\frac{n_s}{n}c^2+\frac{n_n}{n}v_F^2}=v_F\sqrt{(1-t^4) (\frac{c}{v_F})^2       +t^4}
\eeq
within a two-fluid description. Figure 2 shows the obtained behavior of the amplitude of the potentials with this assumption, compared to the experimental
data of Avramenko et al for $Ga$\cite{sound} (2004) for three different values of $v_F/c$. It can be seen that our curves qualitatively and semiquantitavely fit the
observations for reasonable values of $v_F/c$. An even better fit may result from using  values of the superfluid concentration derived from measurement
of the temperature-dependent London penetration depth rather than the two-fluid model temperature dependence assumed here.
We argue that the comparison shown in Fig. 2 provides strong evidence in favor of the greatly enhanced charge rigidity of superconductors
predicted by our theory.

\begin{figure}
 \resizebox{8.5cm}{!}{\includegraphics[width=9cm]{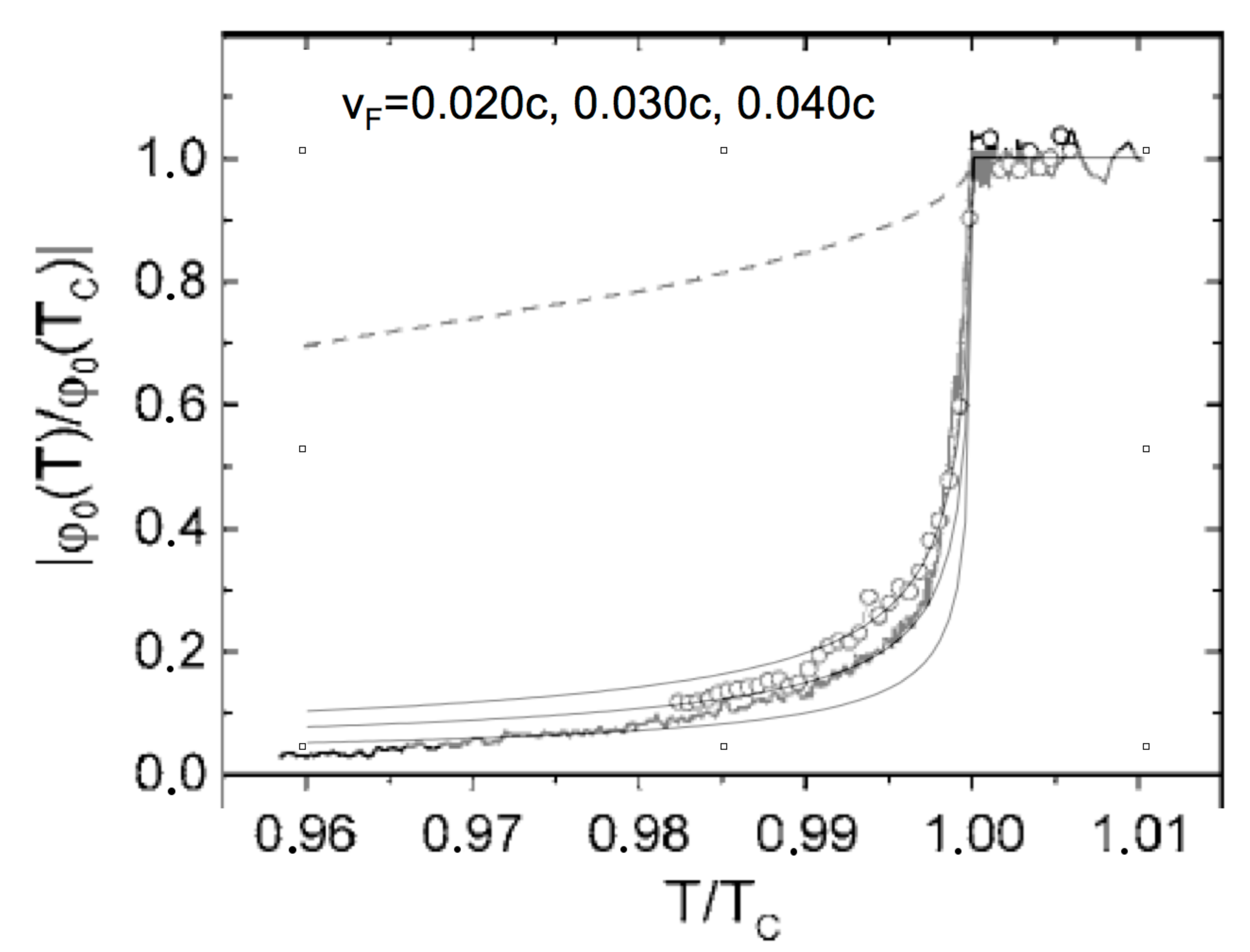}}
 \caption {The decay of the sound signal below $T_c$ measured by Avramenko et al\cite{sound} in Ga for two different values of the sound amplitude
 (jagged solid line (high amplitude) and open circles (low amplitude) and theoretical prediction (Eqs. (30), (31)) for three different values of $v_F/c$ (smooth solid lines).
 The higher line corresponds to the larger $v_F/c$. The dashed line gives the behavior of the Fermi function $f(\Delta)$ (from ref. \cite{sound}) which would be 
 the expected behavior within the conventional theory of superconductivity.}
 \label{figure2}
 \end{figure}

\section{de Heer cluster dipole moments}

In a series of papers, W. de Heer and coworkers\cite{clusters} established that small metallic clusters of $Nb$, $V$ and $Ta$ exhibit a large  electric dipole
moment at low temperatures (of order several Debye for clusters of up to $100$ atoms). Through a variety of measurements they found very strong evidence that the development of the electric dipole moment is
associated with the onset of superconductivity. In contrast, similar clusters of a non-superconducting metal, $Na$, showed essentially no electric
dipole moments\cite{clusters2}. 

Within our theory a superconducting body expels negative charge from the interior to the surface and the resulting charge distribution is rigid. 
The  distribution of electronic charge is determined
by the geometry of the body and can be obtained by numerical solution of the electrostatic equations\cite{ellipsoids}. 
Initially we had hoped\cite{chargeexp} that the inhomogeneous electronic charge distribution predicted by our theory would account for the electric dipole moments observed
by de Heer et al. However our calculations show that
the electronic charge distribution does not exhibit an electric dipole moment even for sample shapes without inversion symmetry\cite{italy}.

However, the distribution of ionic charge in a small cluster is discrete rather than continuous, and this fact is not taken into account in our calculation. Small metallic clusters have irregular shapes\cite{clusters3,cluster4}, in general  with no  inversion symmetry. Generically an electric dipole moment will be generated by the ionic charges determined by the
overall shape of the cluster as well as  by the discrete location of the ions. 
In the normal state, as well as within conventional BCS theory,
metallic or superfluid electrons are extremely efficient at screening  electric fields over a length scale $\lambda_{TF}$,  of order $1\AA$ (eq. (18))
and thus will screen any ionic dipole moment. 
Instead, within our theory the superfluid electrons can only screen electrostatic fields over distances of order the London penetration depth (Eqs. (16), (5)),
typically of order several hundred $\AA$, which is much larger than the linear dimensions of the de Heer clusters (which have up to $\sim$ $100$ atoms and linear
dimensions smaller than $10\AA$). Therefore, we argue that the observation of large unscreened electric dipole moments in metallic clusters of dimensions much
smaller than the London penetration depth is strong evidence in favor of the large  rigidity of the superfluid electron charge distribution predicted by our
dielectric function Eq. (7).

\section{Tao effect}

In a series of papers\cite{tao}, R. Tao and coworkers found that superconducting microparticles in a strong electrostatic field assemble into spherical shapes of
macroscopic dimensions. We have proposed a detailed explanation of this ``Tao effect''\cite{taoeffect}, based on the charge expulsion and resulting electric fields
in the neighborhood of superconducting particles of non-spherical shape predicted by our theory.

However, even without considering the details of our theory, in a more general context it is clear, as pointed out  in the experimental papers\cite{tao}, that
this observation is impossible to explain unless electrostatic fields penetrate the superconducting particles a distance considerably larger than the
Thomas Fermi length. This then $requires$ that the superconducing charge distribution is more rigid than in the normal state where it can  screen
the electric field beyond an $\AA$ or so of the surface.
Thus, we argue that the observation of the Tao effect is also a strong indicator that the charge distribution in superconductors is more rigid than in the normal state.

\section{screening of electrostatic fields}

The electrodynamic equations of  our theory predict\cite{electrodyn}, according to Eq. (16),
 that the superfluid electrons screen applied static electric fields over a distance $\lambda_L$ rather than over 
a Thomas Fermi screening length as predicted by BCS theory\cite{conv3,koyama}.  
In fact, the London brothers themselves  considered electrodynamic equations  for superconductors predicting such behavior in an early version of their theory\cite{londonearly}.
However, shortly thereafter H. London performed an experiment\cite{londonexp} attempting to detect this effect and
didn't find it, after which
the London brothers discarded that version of the theory and adopted the conventional London equations to describe superconductors which do not allow electric fields in
the interior.

We  expect the electric screening length to increase continuosly  from $\lambda_{TF}\sim 1\AA$ to $\lambda_L\sim$ $100$'s of $\AA$ 
as the temperature is lowered from $T_c$ to $0$. In a 2-fluid model description the static dielectric constant at finite temperatures is given by
\beqn
\epsilon(k,0)&=&\epsilon_s(k,0)+\epsilon_n(k,0) -1 \nonumber \\
&=&1+(\frac{1}{\lambda_L(T)^2}+\frac{1}{\lambda_{TF}(T)^2})\frac{1}{k^2}
\eeqn
giving the effective electric screening length $\lambda_E$ as
\beq
\frac{1}{\lambda_E(T)^2}=\frac{1}{\lambda_L^2} \frac{n_s(t)}{n}+ \frac{1}{\lambda_{TF}^2} \frac{n_n(t)}{n}
\eeq
witn $n_s(t)=n(1-t^4)$, $n_n(t)=nt^4$, with $t=T/T_c$. Eq. (33) predicts the temperature dependence of the electric screening length shown in Fig. 3.
Note that only at temperatures well  below $T_c$ does the screening length increase substantially.

In H. London's 1936 experiment\cite{londonexp} he attempted to measure changes in the capacitance of a capacitor with superconducting electrodes of the metal $Hg$
that would result from an increased electric penetration depth. 
His experiment showed no change, from which he concluded that the electric screening length doesn't change in superconductors. However, the lowest temperature reached in H. London's experiment was $T=1.8^oK$, which corresponds to $T/T_c=0.43$ for $Hg$ ($T_c=4.153 ^oK$).
With the sensitivity of his experiment, London could have detected a change in the capacitance corresponding to the screening length increasing above
$\lambda_E\sim 20\AA$. As can be seen in Fig. 3, for $T/T_c\sim0.4$ the screening length would only have increased to about $5\AA$, hence
substantially less than what could have been detected with the sensitivity of that experiment. 

In 1993 Jenks and Testardi attempted to measure the penetration of a static electric field into epitaxial thin films of $YBa_2Cu_3O_{7-x}$\cite{testardi}. 
They reported a large change close to $T_c$, in apparent disagreement with both BCS theory and with our expected behavior shown in Fig. 3.
However, it is not clear that this experiment was free of experimental artifacts, since the results also showed a large change in penetration depth with
temperature $above$ $T_c$, and variations between different films. The experiment has not been repeated, nor are there any other  published reports of
 attempts to measure changes in the electric screening length below the superconducting critical temperature in either high $T_c$ or conventional materials
 to our knowledge.
 
   \begin{figure}
 \resizebox{8.5cm}{!}{\includegraphics[width=9cm]{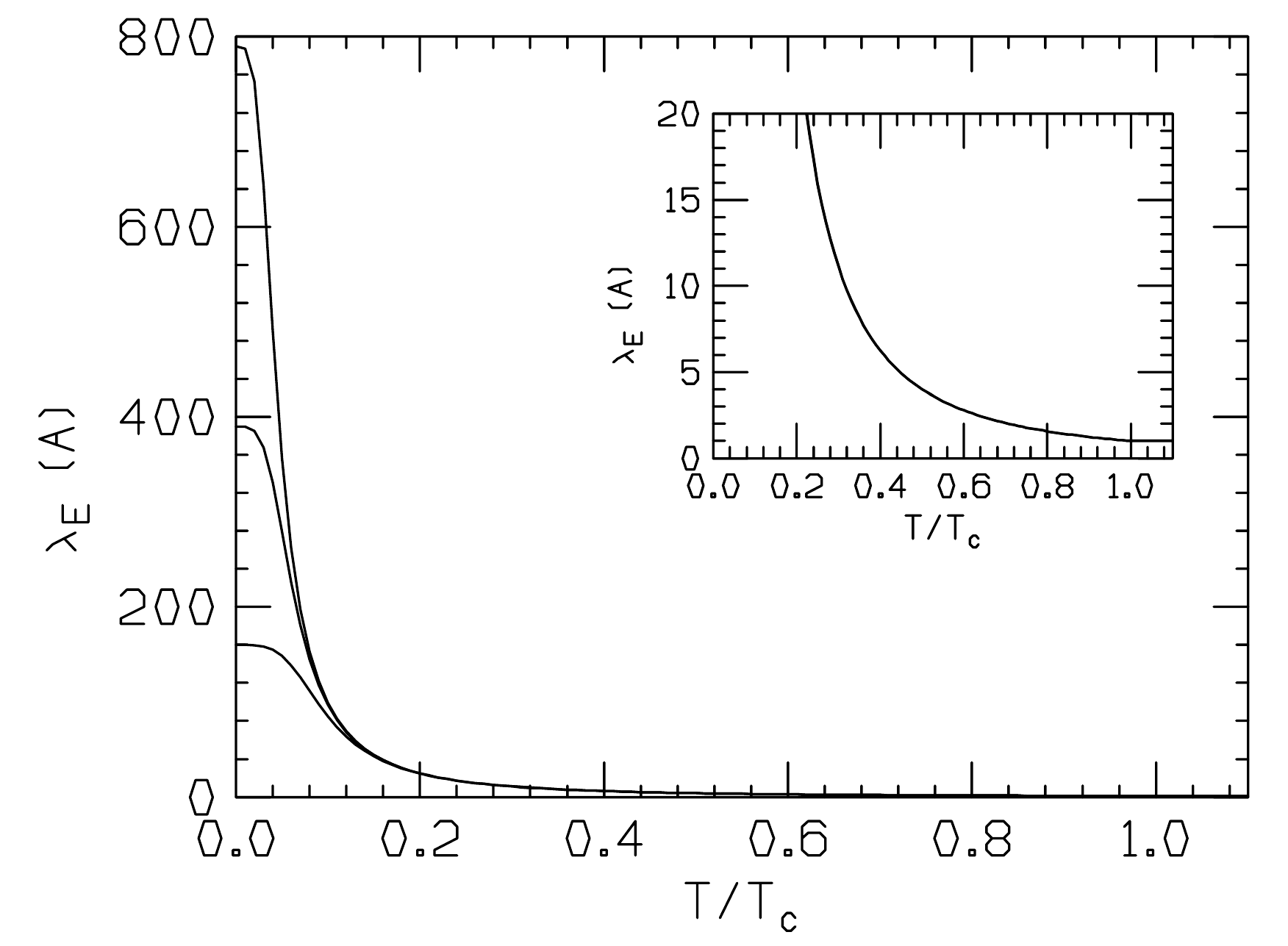}}
 \caption {Screening length for electrostatic field versus temperature for three values of the zero temperature London penetration depth
 ($\lambda_L=790 \AA$, $390 \AA$ and $160 \AA$)
 appropriate for $Hg$, $Nb$ and $Al$ respectively. The inset shows the same data on a different scale.
 $\lambda_{TF}=1\AA$ is assumed.
 }
 \label{figure2}
 \end{figure}

\section{discussion}

The need for reformulation of London electrodynamics arose in our theory from the prediction that negative charge expulsion occurs in the
transition to superconductivity\cite{chargeexp0}, which is a consequence of the microscopic physics of electron-hole asymmetry\cite{undressing2} and incompatible with the conventional
London equations that assume that no electrostatic fields can exist in superconductors. Our reformulation renders  the theory 
relativistically covariant\cite{electrodyn}, unlike conventional London electrodynamics, and   allows for 
a natural and consistent extension of the
electrodynamics equations to the spin sector\cite{electrospin} so as to describe both charge and spin currents, which is necessitated by the
predicted existence of an outward pointing  electric field in the interior of superconductors.

The enhanced charge  rigidity and inability to screen can be seen  also as a natural consequence of several other aspects of the theory. For example, superconductivity in this theory is driven by kinetic energy rather than potential energy lowering\cite{kinenergy,nine}. Thus, in contrast to the normal
metal the superconductor is willing to pay a price in Coulomb potential energy  in order to optimize kinetic energy, which naturally results in its inability to effectively screen electric fields over short distances, a process which is potential-energy driven in the normal metal. 
Kinetic energy lowering is associated with the fact that in the transition to superconductivity  electrons `undress' from the electron-ion interaction,
expand their wavelength and no longer ``see' the discrete ionic potential\cite{eh2}, hence are unable to screen perturbations   on
interatomic distance scales as normal electrons do.

Associated with the much larger screening length is the fact that the compressibility of the superfluid is enormously reduced
compared to the normal metal. This is related to the enhanced quantum pressure of the superfluid compared to the 
normal fluid, which is manifest in superconductors in the negative charge expulsion and in superfluid $^4He$ in the
fountain effect\cite{kinsuperfluids}. It does not mean however that the pressure is increased by the same factor as the rigidity (bulk modulus).
We have for the superconductor (Eq. (15))
\beq
B\equiv\frac{1}{\kappa}=n\frac{\partial P}{\partial n}=m_e c^2 n
\eeq
and integrating we obtain
\beq
P=m_e c^2 (n-n_0)
\eeq
where $n_0$ is an integration constant, in contrast to the normal metal where
\beq
P=
\frac{2}{5} \epsilon_F n .
\eeq
It is natural to conclude that $n-n_0$ is of order $\rho_-/e$ ($\rho_-$ is the excess negative charge density near the surface\cite{electrospin}), the expelled number density, which is smaller than the
superfluid density $n$ by about the same factor ($\sim 10^6$) than the energy $m_e c^2$ is larger than the
Fermi energy $\epsilon_F$\cite{electrospin}. Thus the superfluid pressure in the superconductor is of the same order of magnitude as the
electronic pressure in the normal metal, but its rigidity is enormously enhanced.

How is this compatible with the experimental observation that the compressibility of a solid in the superconducting state is
essentially the same as in the normal state? Clearly in a quasistatic compressibility measurement the ions and the electrons
move together, no charge imbalance is generated and the enhanced rigidity does not manifest itself. It is only in
experiments where the electronic density is locally changing relative to the ionic density that the much larger rigidity will
show up.

 \begin{figure}
 \resizebox{8.5cm}{!}{\includegraphics[width=9cm]{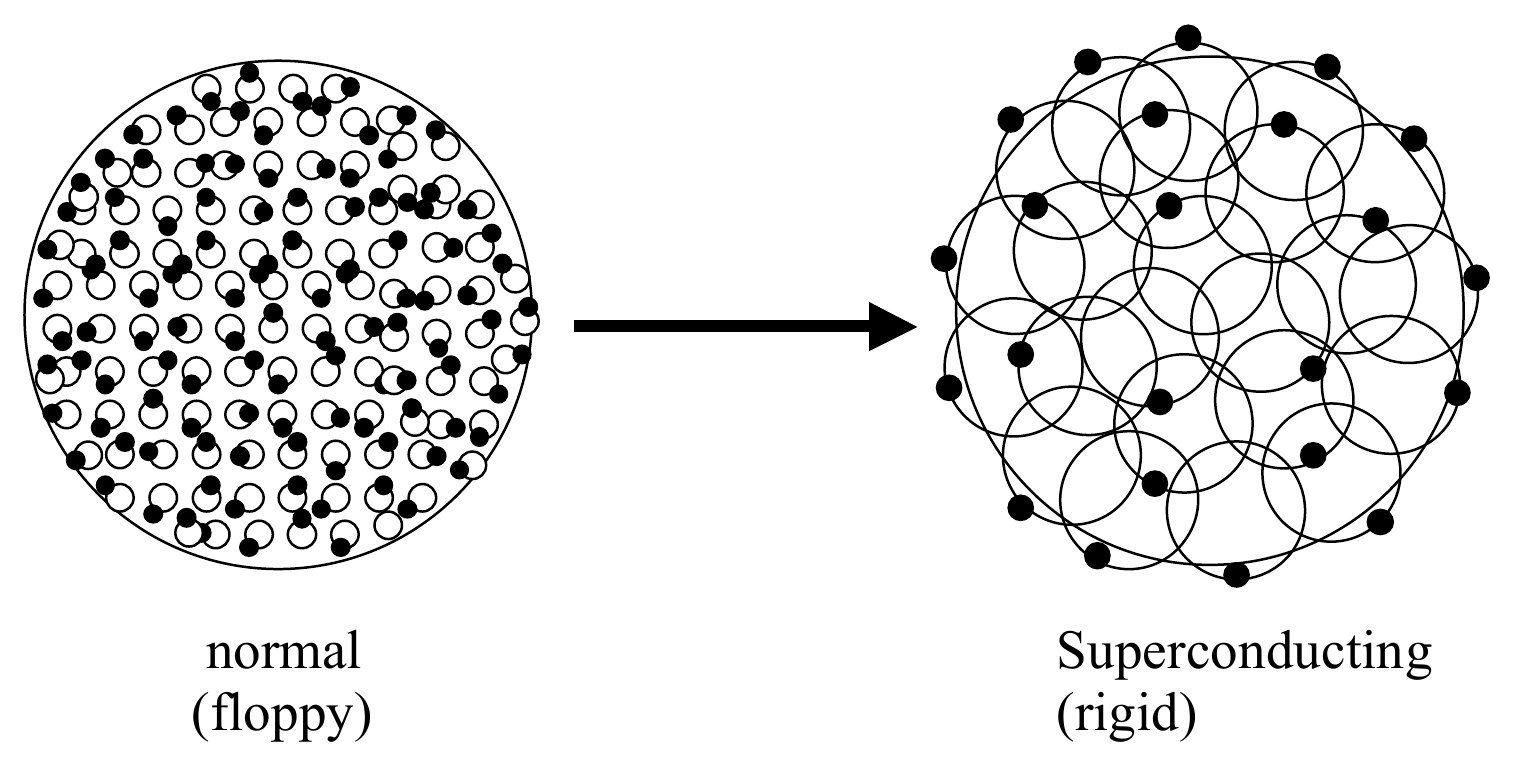}}
 \caption {Electronic orbits expand from radius $k_F^{-1}$ in the normal state (left)  to radius $2\lambda_L$ in the superconducting state (right). 
 This is the origin of the charge expulsion and the charge rigidity over distances of order $\lambda_L$ predicted by the theory.
 The black dots denote the instantaneous position of the electron, i.e. the ``phase'', which is random in the normal state where the orbits are
 non-overlapping and highly correlated
 between different overlapping orbits (i.e. phase coherent) in the
 superconducting state. The orbiting speed is $v_\sigma^0=\hbar/(4m_e\lambda_L)$ in the superconducting state\cite{electrospin}.
 }
 \label{figure2}
 \end{figure}
 
Formally one can write electrodynamics equations for the superconductor where the screening length for electrostatic fields
is $\lambda_L$ but where no charge expulsion occurs, as done by the London brothers themselves    in the early version of their theory\cite{londonearly},
as well as by others thereafter\cite{govaerts,g2}.  Mathematically
the formalism is very appealing but there is no physics behind it, and perhaps for that reason the London brothers
were quick to  discard it  soon thereafter when H. London's experiment seemed to disprove it\cite{londonexp}. For us instead, 
the enhanced charge rigidity and the predicted charge
expulsion are inextricably linked: no enhanced rigidity can take place
without charge expulsion and no charge expulsion can occur without accompanying enhanced rigidity. This is because both phenomena
are a direct  consequence of the fact that electronic orbits expand, driven by kinetic energy lowering, from microscopic non-overlapping orbits of radius
$k_F^{-1}$ to orbits of radius $2\lambda_L$ in the transition to superconductivity\cite{sm}, as shown schematically in Fig. 4. Orbit expansion implies outward motion of negative
charge, and the resulting mesoscopic orbits are highly overlapping which  makes it impossible to create a charge fluctuation over a small distance since the extra electrons would not
have the ability to insert their orbits in the mesh of highly correlated interpenetrating orbits that already exists.

It is often said in the context of the conventional theory of superconductivity 
that the wavefunction of a superconductor is ``rigid'', a concept first introduced by F. London. In the conventional theory, ``rigidity'' refers only to the
response to magnetic perturbations. Instead, our theory extends the property of rigidity of superconductors also to the response to electric perturbations.
Rigidity to both magnetic and electric perturbations originates in the overlapping phase-coherent $2\lambda_L$ orbits depicted in Fig. 4, which also explains
 the macroscopic phase coherence (phase rigidity) of the superconductor: an electron orbiting out of phase would collide with other electrons in
overlapping orbits and pay a high price in Coulomb energy.
And this also explains why the length $\lambda_L$ enters symmetrically in our theory for both magnetic and
 electric phenomena\cite{electrospin}: the $2\lambda_L$ orbits are necessary for the Meissner effect to take place\cite{lastpaper}, as already suspected
 long ago by Smith\cite{smith} and  by Slater\cite{slater}, and the same $2\lambda_L$ orbits determine the electric screening length. 
 The wavefunction for the superconducting
 state has to describe superfluid electrons in $2\lambda_L$ orbits, which BCS theory does not do, if it is to
 describe the ubiquitous  Meissner effect and the experimental consequences of enhanced charge rigidity discussed
 in this paper.
 
The fact that the superfluid wavefunction is rigid with respect to both magnetic and electric perturbations and the resulting new
electrodynamics follow naturally in a relativistic context\cite{londonearly}.
Within Klein-Gordon theory\cite{baym} describing a relativistic scalar wavefunction $\Psi(\vec{r},t)$, the current four-vector 
${\it{J}}=(\vec{J}(\vec{r},t),ic\rho(\vec{r},t))$  for the current and charge densities is given by
\bmath
\beqn
\vec{J}(\vec{r},t)&=&\frac{e}{2m}[\Psi^*(\frac{\hbar}{i}\vec{\nabla}-\frac{e}{c} \vec{A}(\vec{r},t))\Psi+ \nonumber \\
& &
\Psi(-\frac{\hbar}{i}\vec{\nabla}-\frac{e}{c} \vec{A}(\vec{r},t))\Psi^*]
\eeqn
\beqn
\rho(\vec{r},t)&=&\frac{e}{2mc^2}[\Psi^*(i\hbar \frac{\partial}{\partial t} -e \phi(\vec{r},t))\Psi+ \nonumber \\
& &\Psi(-i\hbar \frac{\partial}{\partial t} -e \phi(\vec{r},t))\Psi^*]
\eeqn
\emath
with $\vec{A}$ the magnetic vector potential and $\phi$ the electric potential. In the conventional theory it is said that the Meissner effect results from the fact that
the wavefunction is unaffected by changes in the magnetic vector potential $\vec{A}$. Hence, since $\vec{J}=0$ in the absence of magnetic fields,
eq. (37a) implies that for any value of $\vec{A}$
\beq
\vec{J}(\vec{r},t)=-\frac{n_s e^2}{m_e c}\vec{A}(\vec{r},t)
\eeq
with $n_s=\Psi^*\Psi$, giving rise to the Meissner effect. Extending the argument  it is natural to
assume that the wavefunction $\Psi(\vec{r},t)$ is also unaffected by applied electric fields and by proximity to the boundaries of the sample. If deep in the
interior of the superconductor the charge density is assumed to be a constant $\rho_0$, with associated electric potential
$\phi_0(\vec{r})$ ($\nabla^2 \phi_0=-4\pi\rho_0$), it follows from applying Eq. (37b) to a position deep in the interior and 
another arbitrary position $\vec{r}$ and substracting,    that at any position $\vec{r}$ with or without applied electric fields 
\beq
\rho(\vec{r},t)-\rho_0=-\frac{n_s e^2}{m_e c^2} (\phi(\vec{r},t)-\phi_0(\vec{r}))
\eeq
which is the basic equation of our modified electrodynamic formalism\cite{electrodyn} determining the charge distribution and electric potential in 
superconductors of arbitrary shape.

In summary we have discussed in this paper  six  different   experimental probes  of the enhanced charge rigidity of superconductors predicted by
our theory. Three of them (electron sound, de Heer effect, Tao effect) have already shown
 clear evidence for enhanced charge  rigidity. Another two (bulk and surface plasmons) have not yet been
experimentally tested, and there are a variety of different experimental techniques (EELS, IXS, optical transmission, optical near-field, nanoparticles) that can be
used for that purpose. Finally, the predicted  increased electric screening length below $T_c$ has yielded ambiguous results
so far\cite{londonexp,testardi}. For none of the three observations
that we interpret as arising from   the enhanced charge rigidity predicted by our theory have alternative plausible explanations been
proposed, and they all seem to be incompatible with conventional London-BCS theory. 
For the changes that we predict in the bulk and surface plasmon dispersion relations no other
such predictions have been made in other theoretical frameworks and they are also incompatible with conventional
BCS-London theory, as is the predicted increase in electric screening length at low 
temperatures. It will be interesting to confront the predictions of our theory and of BCS-London theory with future experimental results.

  \acknowledgements{}
  The author is grateful to W. de Heer, R. Tao, J. Fink, D. Basov and R.  Prozorov for stimulating discussions.

\end{document}